\documentclass{sigchi}

\toappear{}
\usepackage{balance}  
\usepackage{graphics} 
\usepackage{times}    
\usepackage{url}      
\usepackage{graphicx}
\usepackage{caption}
\DeclareCaptionType{copyrightbox}
\usepackage{subcaption}
\def\etal{et~al.\,}
\makeatletter
\def\url@leostyle{%
  \@ifundefined{selectfont}{\def\UrlFont{\sf}}{\def\UrlFont{\small\bf\ttfamily}}}
\makeatother
\def\pprw{8.5in}
\def\pprh{11in}

\usepackage[pdftex]{hyperref}
\hypersetup{
pdftitle={Uncovering the Wider Structure of Extreme Right Communities Spanning Popular Online Networks},
pdfauthor={Derek O'Callaghan, Derek Greene, Maura Conway, Joe Carthy, Padraig Cunningham},
pdfsubject={Social network analysis; extreme right; heterogeneous online networks},
pdfkeywords={Social network analysis; extreme right; heterogeneous online networks},
bookmarksnumbered,
pdfstartview={FitH},
colorlinks,
citecolor=black,
filecolor=black,
linkcolor=black,
urlcolor=black,
breaklinks=true,
}

\begin{document}

\title{Uncovering the Wider Structure of Extreme Right Communities Spanning Popular Online Networks}

\numberofauthors{5}
\author{
  \alignauthor Derek O'Callaghan\\
    \affaddr{University College Dublin}\\
    \affaddr{Dublin 4, Ireland}\\
    \email{derek.ocallaghan@ucd.ie}\\
  \alignauthor Derek Greene\\
     \affaddr{University College Dublin}\\
    \affaddr{Dublin 4, Ireland}\\
    \email{derek.greene@ucd.ie}\\
   \alignauthor Maura Conway\\
   \affaddr{Dublin City University}\\
    \affaddr{Dublin 9, Ireland}\\
    \email{maura.conway@dcu.ie}
\and
   \alignauthor Joe Carthy\\
   \affaddr{University College Dublin}\\
    \affaddr{Dublin 4, Ireland}\\
    \email{joe.carthy@ucd.ie}\\
   \alignauthor P\'{a}draig Cunningham\\
   \affaddr{University College Dublin}\\
    \affaddr{Dublin 4, Ireland}\\
    \email{padraig.cunningham@ucd.ie}\\
}


\maketitle

\begin{abstract}
Recent years have seen increased interest in the online presence of extreme right groups. Although originally composed of dedicated
websites, the online extreme right milieu now spans multiple networks, including popular social media platforms such as Twitter, Facebook
and YouTube. Ideally therefore, any contemporary analysis of online extreme right activity requires the consideration of
multiple data sources, rather than being restricted to a single platform. We investigate the potential for Twitter to act as one possible
gateway to communities within the wider online network of the extreme right, given its facility for the dissemination of content. A strategy
for representing heterogeneous network data with a single homogeneous network for the purpose of community detection is presented, where these
inherently dynamic communities are tracked over time. We use this strategy to discover and analyze persistent English and German language
extreme right communities.
\end{abstract}




\section{Introduction}

The online presence of the extreme right has come under greater scrutiny in recent years, particularly given events such as the Oslo
killings by Anders Behring Breivik in July 2011, Wade Michael Page's attack on a Sikh temple in Wisconsin in August 2012 and the uncovering
in November 2011 of the German National Socialist Underground (NSU)'s multi-year murder campaign. While initially the extreme right relied
upon dedicated websites for the dissemination of their hate content \cite{Burris2000}, they have since incorporated the use of social media
platforms for community formation around variants of extreme right ideology \cite{ISDPreventingCounteringExtremism2012}. As Twitter's
facility for the dissemination of content has been established \cite{Bakshy:2011:EIQ:1935826.1935845}, our previous work
\cite{OCallaghan2012} provided an exploratory analysis of its use by extreme right groups. Here we extend this analysis to investigate its
potential to act as one possible gateway to communities located within the online extreme right milieu, which spans multiple
platforms such as Twitter, Facebook and YouTube in addition to other websites. Our analysis is focused on two case studies, involving the investigation of
English and German language online communities.

We infer network representations of heterogeneous nodes and edges to capture the relations between four different extreme right online
entities: (a) Twitter accounts, (b) YouTube channels, (c) Facebook profiles and (d) all other websites. The inclusion of such platforms
allows us to identify extreme right communities that would otherwise not be apparent when considering Twitter data alone. As online extreme
right networks are inherently dynamic, we are also interested in tracking the evolution of the associated groups over an extended period of
time. Following work by Lehmann \etal~\cite{lehmann08biclique}, we apply the community finding process to a single network representation
where all node types are treated as peers. When interpreting the output of the process, the use of heterogeneous data provides us with a
rich insight into the composition of the communities, and serves to differentiate between them.

We initially provide a description of related work on the online activities of extremist groups, along with discussion of relevant network
analysis methods. The generation of two data sets using English and German language tweets is then discussed. Next, we describe the
methodology used to derive network representations and subsequently track detected communities. This is followed by an analysis of selected
persistent dynamic communities found in both data sets, where we discuss the merits of the network representation and provide
characterizations of the membership composition in terms of the extent to which they span multiple online platforms. Finally, the overall
conclusions are discussed, and some suggestions for future work are made.

\section{Related Work}
\label{relatedwork}

\subsection{Online Extremism}
A number of studies have investigated the online presence of extreme right groups.
An early example is that of Burris~\etal \cite{Burris2000}, where social network analysis 
of a white supremacist website network 
found evidence of decentralization, with little division along doctrinal lines.
Gerstenfeld~\etal \cite{GerstenfeldHateOnline2003} analyzed the content of various extremist websites, 
and described  
the potential for
forging a stronger sense of community between geographically isolated groups. 
Chau and Xu \cite{Chau:2007:MCR:1222244.1222622} studied networks built from users contributing to
hate group and racist blogs.
Caiani and Wagemann \cite{doi:10.1080/13691180802158482} analyzed the website networks of German and Italian extreme right groups.

More recent work has included the study of online social media usage by the extreme right. Bartlett~\etal \cite{BartlettDigitalPopulism2011}
performed a survey of European populist party and group supporters on Facebook.
Baldauf~\etal
\cite{ZwischenPropaganda2011} investigated the use of Facebook by the German extreme right, 
exploring 
the strategies employed for promoting associated ideology 
along with
the use of certain themes designed to
attract new unsuspecting followers, such as stated support for freedom of speech and opposition to child sex abuse. 
Goodwin and Ramalingam's
\cite{GoodwinRadicalRight2012} analysis of the radical right in Europe discussed the importance of online social media, and separately found
a shortage of research on non-electoral forms of right-wing extremism.
A set of reports commissioned by the Swedish Ministry of Justice and the Institute for Strategic Dialogue
\cite{ISDPreventingCounteringExtremism2012} 
analyzed 
the use of social media such as Facebook and YouTube by a number of European extreme
right groups. 

These related studies suggest that
communities within the online extreme right milieu span multiple networks, which provides a motivation for the current work to analyze their
structure and temporal persistence, while also investigating the potential for popular social media platforms such as Twitter to act as 
gateways to this activity.

\subsection{Network Analysis}

An appropriate network representation is required in order to analyze the multi-network online activity of the extreme right. The
\emph{inference} and \emph{relevance} problems discussed by De Choudhary \etal~\cite{DeChoudhury:2010:IRS:1772690.1772722} are applicable
here, where the former is attributed to the fact that ``real'' social ties are not directly observable and must be inferred from
observations of events (in our case, tweets from identified extreme right accounts on Twitter), while the lack of one ``true'' social
network is associated with the latter, due to the potential existence of multiple networks each corresponding to a different definition of a
tie. Separate \emph{incompleteness} issues also exist given the nature of the extreme right. These issues are similar to those encountered
by Sparrow~\cite{Sparrow1991251} and Krebs~\cite{Krebs_2002} in their respective studies of criminal and terrorist networks, where
references are made to the inevitability of missing nodes and links, the difficulty in deciding who to include and who not to include, and
the dynamic quality of the networks. As our current work uses Twitter as a starting point for analyzing online extreme
right activity, the notion of missing nodes and links is realized with the unavailability of older tweet data, relevant Twitter accounts that have been
suspended, profiles that are not publicly accessible and extreme right entities that do not maintain a presence on Twitter.

Our proposed representation of heterogeneous node types with a single homogeneous network is also motivated in part by previous work in
cluster analysis and community finding, where nodes with different semantics have been treated equally during the clustering phase, and
subsequently treated separately to support the interpretation of the resulting clusters. For instance, Dhillon~\cite{dhillon01cocluster}
produced a bipartite spectral embedding of words and documents, which allowed both to be clustered simultaneously. Lehmann
\etal~\cite{lehmann08biclique} proposed an extension of the well-known clique percolation method for community finding to identify
communities of nodes in bipartite graphs. The use of multiple node types avoids the loss of important structural information, which can
occur when performing a one-mode projection of a graph.

For the analysis of communities in temporal networks, a  variety of approaches have been proposed. Greene \etal
\cite{Greene:2010:TEC:1900720.1900760} described a general model for tracking communities in dynamic networks. Communities are identified on
each individual \emph{time step} network, which provides a snapshot of the data at a particular interval. These communities are then matched
together to form timelines, representing long-lived dynamic communities that exist over multiple successive steps. This work was extended in
\cite{greene10dynak} to address the problem of clustering dynamic bipartite networks, where a clustering is produced on a spectral embedding
of both types of nodes simultaneously at each time step before matching.

\section{Data}
\label{data}

In our previous analysis \cite{OCallaghan2012}, we collected a number of Twitter data sets, each associated with a particular country, to
facilitate the analysis of contemporary activity by extreme right groups. One of our findings was the influence of linguistic proximity on
relationships within the detected extreme right communities, in particular, the use of the English language. Given this, we merged two of
our data sets associated with the USA and the United Kingdom respectively. This data set, along with that associated with German extreme
right Twitter accounts, are the starting points for this current work. Both data sets were extended to include additional accounts that were
identified as relevant, using criteria such as profiles containing extreme right symbols or references to known
groups; follower relationships with known relevant accounts; similar Facebook profiles/YouTube channels; extreme right media accounts; a
propensity to share links to known extreme right websites. Similar criteria were employed in earlier studies
\cite{ZwischenPropaganda2011,doi:10.1080/13691180802158482}, and further details can be found in our previous analysis
\cite{OCallaghan2012}. In total, 1,267 English language and 430 German language accounts were identified. Profile data including followers,
friends, tweets and list memberships were retrieved for these accounts, as limited by the Twitter API restrictions effective at the time,
between June 2012 and November 2012.

We are interested in the potential for Twitter to act as one possible gateway to online extreme right activity through the dissemination of
links to external websites; for example, dedicated websites managed by particular groups, content sharing websites such as YouTube, or mainstream
websites hosting content that could be used to promote associated concerns. Accordingly, the account tweet content is analyzed to extract
all external URLs. YouTube URLs are further processed, where profile data is retrieved for any detected video and channel (account)
identifiers. Similarly, profile data is retrieved for any user, group and page identifiers detected within Facebook URLs. In both cases, we
are restricted to data that is publicly accessible. We also extract Twitter account identifiers from any URLs that point to content such as
photos or other images that are hosted directly on Twitter. Regarding interactions within Twitter itself, we extract all \emph{mention} and
\emph{retweet} events between the identified accounts. Statistics of the final data sets used in the analysis can be found in
Table~\ref{tab:datasets}.

\begin{table*}[!t]
\begin{center}
\begin{tabular}{| l | c | c | c | c | c | c |}
\hline 
\emph{Data set} & \emph{Tweets} & \emph{Mentions} & \emph{Retweets} & \emph{All URLs} & \emph{YouTube URLs} & \emph{Facebook URLs}
\\
\hline \hline 
English language \hspace{1em} & 1,517,339 & 539,181 & 162,042 & 972,444 & 71,049 & 23,007 \\ \hline
German language & 54,873 & 14,263 & 11,028 & 88,309 & 3,868 & 6,924 \\ \hline
\end{tabular}
\end{center}
\caption{Data set statistics for the period of June 1st, 2012 to November 16th, 2012.}
\label{tab:datasets}
\end{table*}

\section{Methodology}
\label{methodology}

\subsection{Network Representation}

In this work, we use a single network representation consisting of four node types: (a) Twitter accounts, (b) YouTube channels, (c) Facebook
profiles and (d) other websites, where we refer to the latter three as \emph{external} nodes (with respect to Twitter). Heterogeneous edges
are also employed to reflect the variety of possible interactions; (a) a Twitter account mentioning another Twitter account, (b) a Twitter
account retweeting another Twitter account and (c) a Twitter account linking to an external node through the inclusion of a corresponding
URL in a tweet. For the purpose of this analysis, no distinction is made between mention or retweet events. In a similar approach to the
ingredient complement network used by Teng~\etal~\cite{Teng:2012:RRU:2380718.2380757}, we infer edges between external nodes whose URLs have
appeared in tweets from the same Twitter account, in order to further tackle the issue of data incompleteness. At this point, we filter any
YouTube and website nodes whose URLs have appeared in tweets from a single account only. As the number of unique profiles found in tweets
containing Facebook URLs tends to be lower, these nodes are not filtered. We do not filter external nodes that could be considered as
mainstream (for example, newspapers or TV channels), as we are also interested in references to such entities by extreme right accounts. All
edges are treated as undirected, with raw weights based on the frequency of the edge event in question. These weights are normalized with
the use of pointwise mutual information (PMI) as suggested by Teng~\etal~\cite{Teng:2012:RRU:2380718.2380757}, where we extend this process
for heterogeneous node pairs $(a,b)$ as follows:
\begin{equation*}
\textrm{PMI(a,b)} = log\left(1 + \frac{p(a,b)}{p(a)p(b)}\right)
\label{eqn:pmi}
\end{equation*}
For Twitter-Twitter mentions$\mid$retweets edges, the PMI is the probability that two Twitter accounts $a$ and $b$ have a
mention$\mid$retweet relationship against the probability that their respective mentions$\mid$retweets are separate:
\begin{equation*}
p(a,b) = \frac{\# \textrm{ of mentions$\mid$retweets between } a \textrm{ and } b}{\# \textrm{ of mentions$\mid$retweets}}
\end{equation*}
\begin{equation*}
p(a) = \frac{\# \textrm{ of mentions$\mid$retweets featuring } a}{\# \textrm{ of mentions$\mid$retweets}}
\end{equation*}
\begin{equation*}
p(b) = \frac{\# \textrm{ of mentions$\mid$retweets featuring } b}{\# \textrm{ of mentions$\mid$retweets}}
\end{equation*}
For Twitter-external URL edges, the PMI is the probability that a URL external entity $b$ has been tweeted by a particular Twitter account $a$
against the probability that their respective URL tweet occurrences are separate:
\begin{equation*}
p(a,b) = \frac{\# \textrm{ of tweets from } a \textrm{ with URL of } b}{\# \textrm{ of tweets with URLs}}
\end{equation*}
\begin{equation*}
p(a) = \frac{\# \textrm{ of tweets from } a \textrm{ with URLs}}{\# \textrm{ of tweets with URLs}}
\end{equation*}
\begin{equation*}
p(b) = \frac{\# \textrm{ of tweets with URL of } b}{\# \textrm{ of tweets with URLs}}
\end{equation*}
For inferred external-external edges, the PMI is the probability that two URL external entities $a$ and $b$ have been tweeted by a
particular Twitter account against the probability that they have been tweeted by separate accounts:
\begin{equation*}
p(a,b) = \frac{\# \textrm{ of accounts that tweeted URLs of both } a \textrm{ and } b}{\# \textrm{ of accounts that tweeted URLs}}
\end{equation*}
\begin{equation*}
p(a) = \frac{\# \textrm{ of accounts that tweeted URL of } a}{\# \textrm{ of accounts that tweeted URLs}}
\end{equation*}
\begin{equation*}
p(b) = \frac{\# \textrm{ of accounts that tweeted URL of } b}{\# \textrm{ of accounts that tweeted URLs}}
\end{equation*}

The addition of inferred edges between external nodes can occasionally lead to dense networks, particularly with the inclusion of accounts
that are prolific tweeters. As the PMI weights for these edges tend to be normally distributed, we filter any such edges with PMI $< \mu +
2\sigma$. All other edges are retained.

\subsection{Community Detection}

As with our previous analysis \cite{OCallaghan2012}, we use our variant \cite{greene12userlists} of the Lancichinetti and Fortunato
consensus clustering method \cite{lanc12consensus} to generate a set of stable consensus communities from an inferred network instance,
based on 100 runs of the OSLOM algorithm \cite{lancichinetti11oslom}. As before, we selected a value of $0.5$ for the threshold parameter
$\tau$ used with the consensus method, as a compromise between node retention and more stable communities. Although we previously detected
communities within a static graph representation, these networks are inherently dynamic, and so we are interested in the topology of these
communities over an extended period of time. Using the method of Greene~\etal~\cite{Greene:2010:TEC:1900720.1900760}, we represent this
dynamic network as a set of $l$ time step networks $\{{g_1, \dots, g_l}\}$, employing a sliding window approach to generate snapshots of the
nodes and edges in the overall network at successive intervals.

\begin{figure}[!b]
	\centering
    \includegraphics[width=\linewidth]{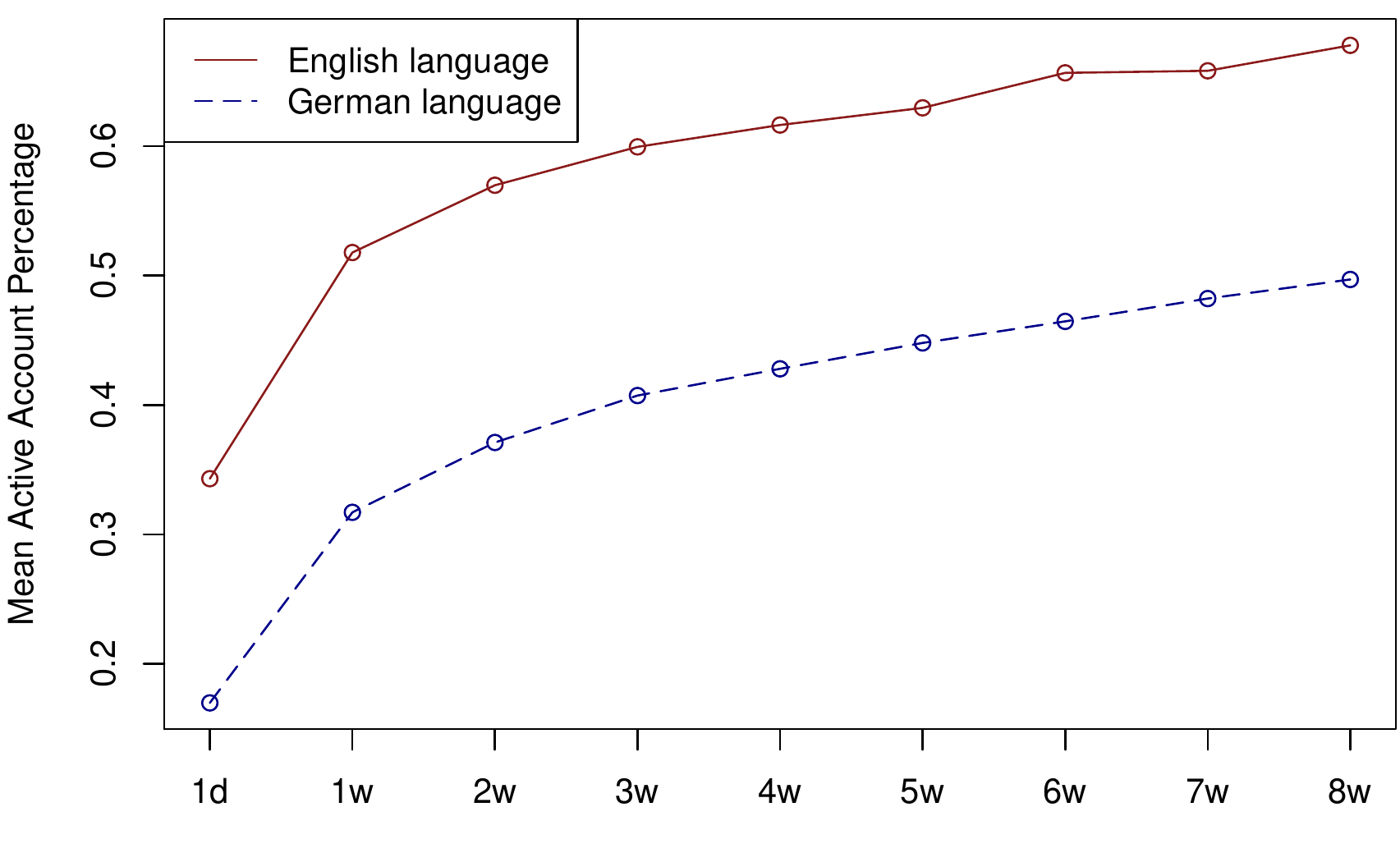}
	\caption{Mean percentage of Twitter accounts active at increasing time scales (1 day up to 8 weeks) for English and German language data sets, between June 2012 and November 2012.}
	\label{fig:timescaleactivity}
\end{figure}

The selection of the sliding window size has a direct impact on the topology of the generated network snapshots. In their discussion of
dynamic proximity networks, Clauset and Eagle~\cite{Clauset2007} claim that dynamics are a multi-scale phenomenon, with variation taking
place at several distinct time scales. However, they also suggest that a natural time scale exists with which important temporal variation
is preserved. In this analysis, we consider this natural time scale to be related to the activity of the identified extreme right Twitter
accounts. We calculated the mean percentage of accounts active in each data set within a series of increasing time scales ranging from one
day up to a duration of eight weeks; a plot of these values can be seen in Figure~\ref{fig:timescaleactivity}. Although the accounts in the
English language data set seem to be more active than those of the German language data set, the relative activity trends are highly
similar. In both cases, there is a sharp increase in the percentage of active accounts between the daily and weekly time scales, with a
lesser increase between the weekly and bi-weekly scales. At this point, the rate of increase slows down, which suggests that using a window
size of two weeks is appropriate for meaningful analysis rather than a larger size whereby the corresponding snapshot networks become
increasingly static. When using non-overlapping sliding windows, we found that the generated step networks exhibited a certain amount of
volatility, potentially due to data incompleteness or a variance in Twitter usage patterns in different countries, for example, the use of
Twitter tends to be more prevalent in the UK and the USA than in other countries such as Germany. We use overlapping sliding windows in an
attempt to smooth this volatility, where the overlap is a period of one week.

Another factor that can influence volatility is related to nodes that are intermittently present in the step networks.
For
example, an offline extreme right event may lead to increased activity from certain Twitter accounts or references to external media nodes.
As a dynamic community timeline is created by matching step communities to existing dynamic communities over successive steps, we address
this issue as follows:

\begin{enumerate}
    \item Rather than solely relying on the most recent observation (referred to as the \emph{front}) in an existing dynamic community
    timeline for comparison with the set of step communities at each step, we consider all historical step communities in a timeline as
    comparison candidates. This permits the consideration of those nodes that are intermittently active.
    \item We use a modified version of the Jaccard coefficient for binary sets to calculate the similarity between a step community
    $C_{ta}$ and a single candidate historical step community $D_{tb}$ belonging to an existing dynamic community timeline, based on the
    \emph{undirected cluster representativeness} metric suggested by Bourqui~\etal~\cite{Bourqui:2009:DSC:1602240.1602667}:
    \begin{equation*}
    \textrm{sim}(C_{ta}, D_{tb}) = \sqrt{\frac{|C_{ta} \cap D_{tb}|}{|C_{ta}|} \times \frac{|C_{ta} \cap D_{tb}|}{|D_{tb}|}}
    \end{equation*}
    \item The overall similarity of a step community with the entire set of candidate communities belonging to an existing dynamic community
    timeline is calculated as an exponential weighted moving average of the individual similarities, with a decay factor $\alpha = 0.5$. A
    match occurs if the overall similarity $\ge 0.25$. This is in line with the range of threshold values used in the original Greene~\etal method evaluation
    \cite{Greene:2010:TEC:1900720.1900760}.
\end{enumerate}

\section{Analysis}

\begin{figure*}[!t]
	\begin{center}
        \begin{subfigure}[b]{0.95\textwidth}
                \centering
                \includegraphics[width=0.93\linewidth]{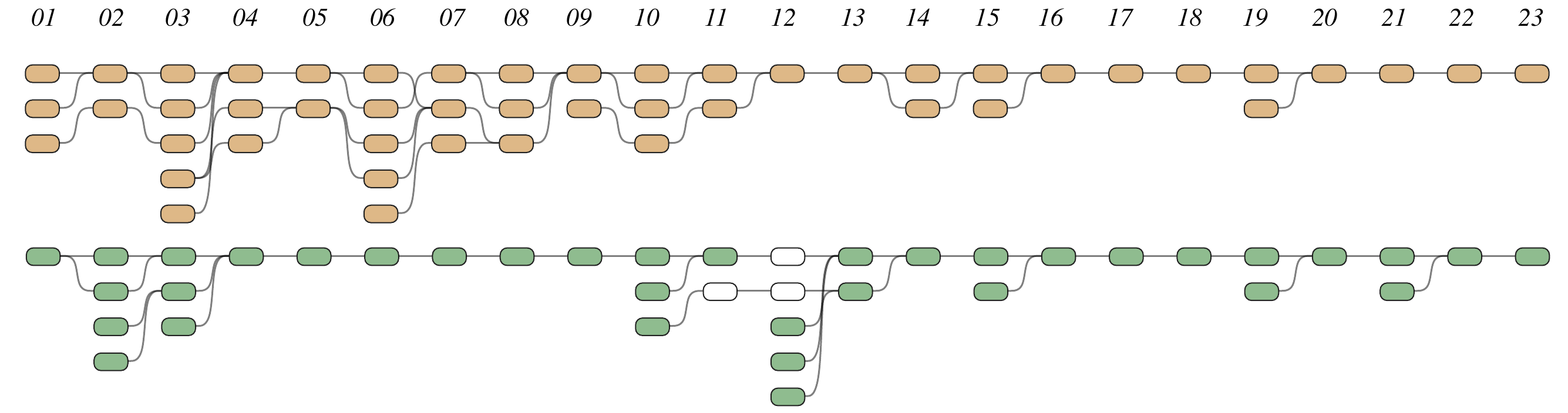}
                \vskip 0.2em
                \caption{EDL (top) and BNP (bottom) communities, containing an average of $14.9\%$ of the nodes in each step network
                ($\sigma=2\%$)}
                \label{fig:etimeline}
        \end{subfigure}
		\vskip 0.8em
        \begin{subfigure}[b]{0.95\textwidth}
                \centering
                \includegraphics[width=0.93\linewidth]{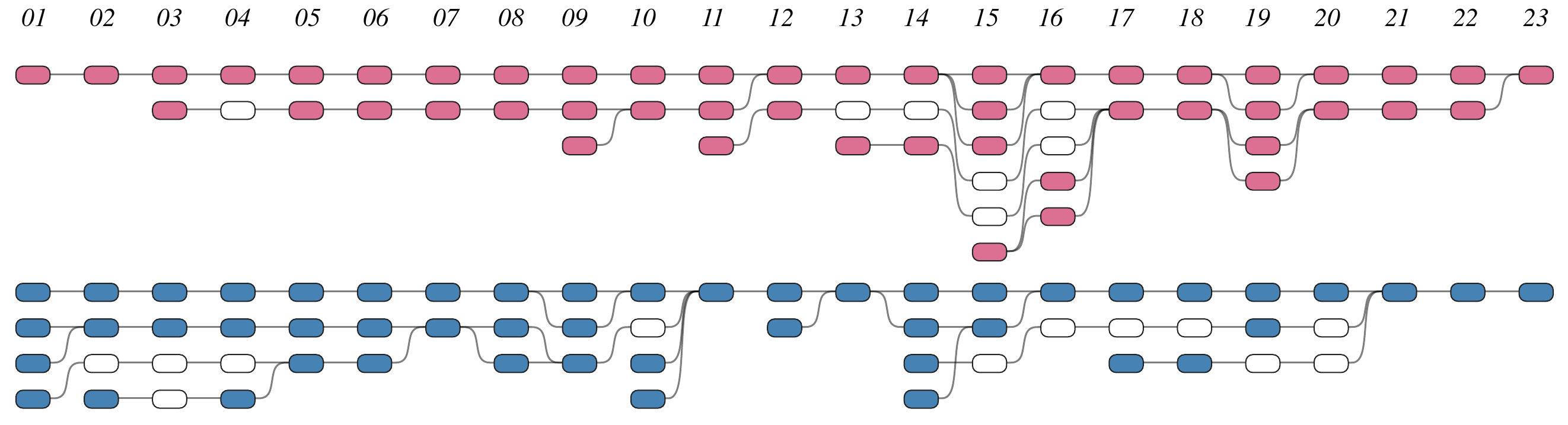}
                \caption{German non-electoral (top) and NPD (bottom) communities, containing an average of $50\%$ of the nodes in each step network
                ($\sigma=7\%$)}
                \label{fig:gtimeline}
        \end{subfigure}
	\end{center}
	\vskip -0.3em
	\caption{Timelines of selected persistent dynamic communities for the (a) English and (b) German language datasets, from June 2012 (step
	$t=1$) to mid-November 2012 (step $t=23$). Each step corresponds to a two week window, with one week overlaps. White rectangles indicate that the
	associated dynamic community was not matched in the corresponding step.}
	\label{fig:timelines}
\end{figure*}

\subsection{Persistent Dynamic Communities}
The method described in the previous section was applied to our two case study data sets, on data collected from June 1st, 2012 to November
16th, 2012. Using a sliding window length of two weeks with a one week overlap resulted in twenty-three time steps. A network representation
was derived for each of these steps, and the dynamic community matching process was then applied to these step networks. In our analysis, we
are particularly interested in \emph{persistent dynamic communities}, where a community is defined as persistent if it has members active in
all time steps. The total number of such dynamic communities found in the English language and German language data sets were fifteen and
four respectively. We have selected two persistent communities from each data set for detailed analysis, and timeline diagrams of the
constituent step communities can be found in Figures \ref{fig:etimeline} and \ref{fig:gtimeline}.

From an initial inspection of these timeline diagrams, it can be seen that these dynamic communities exhibit a certain amount of volatility,
with numerous \textit{merge} and \textit{split} life-cycle events occurring.  As in \cite{Greene:2010:TEC:1900720.1900760}, a merge occurs
if multiple existing dynamic communities are matched to a single step community at time $t$. A split occurs if a single dynamic community is
matched to multiple step communities at time $t$. On occasion, we have found that an existing dynamic community may be a candidate for both
a merge and a split event at a particular time $t$. In this scenario, we give precedence to merge events and a new dynamic community is
created for the step community that generates the split. Both timeline diagrams also include branches of step communities that are distinct
upon initial creation, but become merged in a subsequent step.

An analysis of the nodes in the first English language community in Figure \ref{fig:etimeline} shows that it is associated with the
\textit{English Defence League} (EDL), a street protest movement opposed to the alleged spread of radical Islamism within the UK. This
community also contains nodes associated with the \textit{British Freedom} party, a splinter group from the \textit{British National Party}
(BNP) that formed an alliance with the EDL in 2012 \cite{GoodwinRadicalRight2012,MulhallExtremisBeginnersGuide2012}. We observe the presence
of \textit{Casuals United}, a protest group formed from an alliance of football hooligans, also linked with the EDL. The second English
language community is primarily associated with the BNP, with nodes from other groups such as \textit{Combined ex-Forces} (CxF),
\textit{Infidels} and \textit{The British Resistance} also present \cite{ISDPreventingCounteringExtremism2012}. Other notable persistent
dynamic communities from this data set include two white power/national socialist communities largely consisting of North American and South
African nodes respectively.

The first of the selected German language communities shown in Figure \ref{fig:gtimeline} appears to be associated with a variety of
non-electoral groups. These include \textit{au{\ss}erparlamentarischer Widerstand} (non-parliamentary resistance) entities, \textit{Freies
Netz} (neo-Nazi collectives), along with various ``information/news portals'' \cite{NGNPortals2012}. Two organizations banned by the German
authorities in 2012, namely \textit{Spreelichter} \cite{ZeitSpreelichterBan2012} and \textit{Besseres Hannover}
\cite{ZeitBesseresHannoverBan2012}, are also present. The second German language community is electoral in nature, associated with the
\textit{Nationaldemokratische Partei Deutschlands} (National Democratic Party of Germany - NPD), and includes nodes representing regional
NPD offices and individual politicians. Although this community appears to be associated with regions across Germany, a smaller persistent
NPD community, localized to the federal state of Th\"{u}ringen can also be observed. This smaller community contains a mixture of electoral
(NPD) and non-electoral nodes.

\begin{figure}
	\begin{center}
        \hskip -3.0em
        \begin{subfigure}[b]{0.2\textwidth}
                \centering
                \includegraphics[scale=1.0]{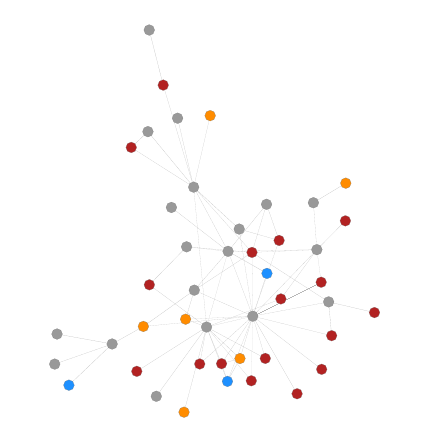}
                \caption{All nodes}
                \label{fig:bnpa}
        \end{subfigure}
       	\quad
		\begin{subfigure}[b]{0.2\textwidth}
                \includegraphics[scale=1.0]{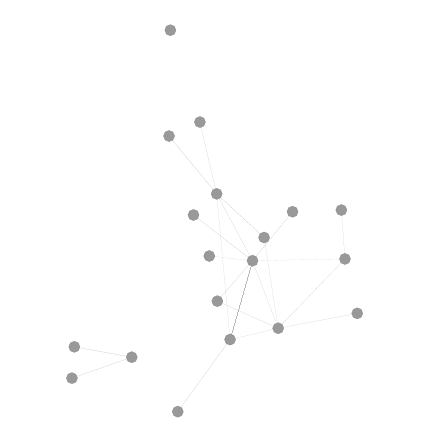}
                \caption{Twitter nodes}
                \label{fig:bnpb}
        \end{subfigure}    
	\end{center}
	\caption{BNP community. Filtering results in central node loss and disconnection (grey=Twitter, blue=Facebook, orange=YouTube, red=other websites).}
	\label{fig:bnpgraphs}
\end{figure}

\subsection{Community Structure}

In this section, we discuss the merits of using a single network representation of heterogeneous node types in our analysis of extreme right
communities. As the online presence of these groups extends beyond individual networks such as Twitter or Facebook, the proposed
representation permits the structure of these wider communities to be revealed, which would otherwise not be evident if analysis was
restricted to a single network. 

\begin{figure}[!b]
	\begin{center}
        \hskip -3.0em
        \begin{subfigure}[b]{0.2\textwidth}
                \centering
                \includegraphics[scale=1.0]{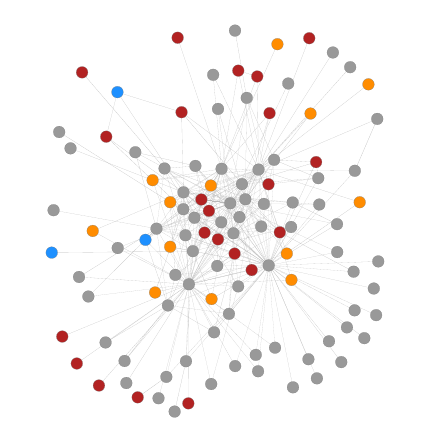}
                \caption{All nodes}
                \label{fig:edla}
        \end{subfigure}
       	\qquad 
		\begin{subfigure}[b]{0.2\textwidth}
                \centering
                \includegraphics[scale=1.0]{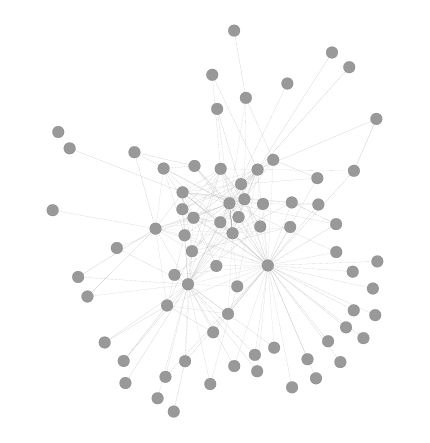}
                \caption{Twitter nodes}
                \label{fig:edlb}
        \end{subfigure}    
	\end{center}
	\caption{EDL community. Filtering results in central node loss (grey=Twitter, blue=Facebook, orange=YouTube,
	red=other websites).}
	\label{fig:edlgraphs}
\end{figure}

We illustrate this by visualizing individual step communities belonging to the persistent dynamic communities selected from the data sets,
as described in the previous section. Network diagrams of these communities were created with Gephi \cite{bastian09gephi}, using the Yifan
Hu layout. Figure~\ref{fig:bnpgraphs} contains diagrams for a BNP step community, with Figure~\ref{fig:bnpa} presenting the community with
all nodes, while non-Twitter nodes have been filtered in Figure~\ref{fig:bnpb} (42\% of nodes and 28\% of edges are retained), resulting in
the loss of important central nodes such as the official BNP website and YouTube channel. The contrast between Figures \ref{fig:bnpa} and
\ref{fig:bnpb} illustrates the important linking role played by non-Twitter nodes, as the observable network is disconnected when they are
not considered. In addition, the use of heterogeneous nodes enables us to immediately identify this community as being associated with the
BNP. Similar diagrams for an EDL step community can be seen in Figure~\ref{fig:edlgraphs}, with 66\% of nodes and 56\% of edges retained
with filtering (Figure~\ref{fig:edlb}). Although disconnection is not introduced, it nevertheless results in the loss of central nodes such
as the official EDL and British Freedom websites.

\begin{figure}[!h]
	\begin{center}
        \hskip -3.0em
        \begin{subfigure}[b]{0.2\textwidth}
                \centering
                \includegraphics[scale=1.0]{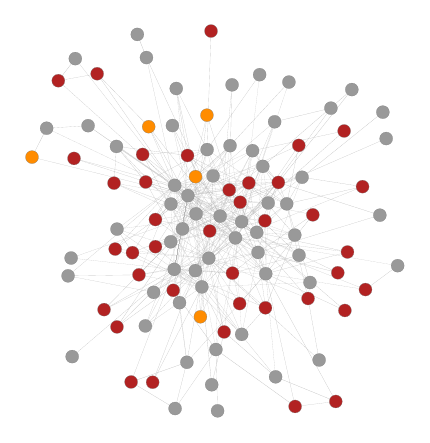}
                \caption{All nodes}
                \label{fig:rada}
        \end{subfigure}
       	\qquad 
		\begin{subfigure}[b]{0.2\textwidth}
                \centering
                \includegraphics[scale=1.0]{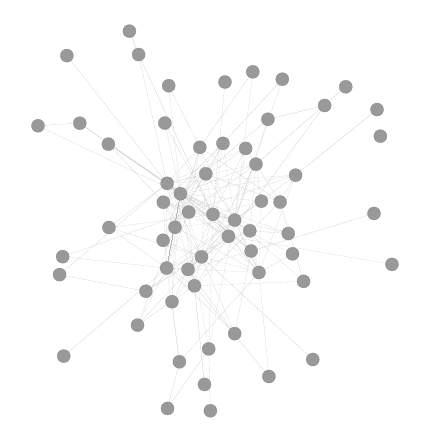}
                \caption{Twitter nodes}
                \label{fig:radb}
        \end{subfigure}    
	\end{center}
	\caption{German non-electoral community. Filtering results in central node loss (grey=Twitter, orange=YouTube, red=other websites).}
	\label{fig:radgraphs}
\end{figure}

\begin{figure}[!h]
	\begin{center}
        \hskip -3.5em
        \begin{subfigure}[b]{0.2\textwidth}
                \centering
                \includegraphics[scale=1.0]{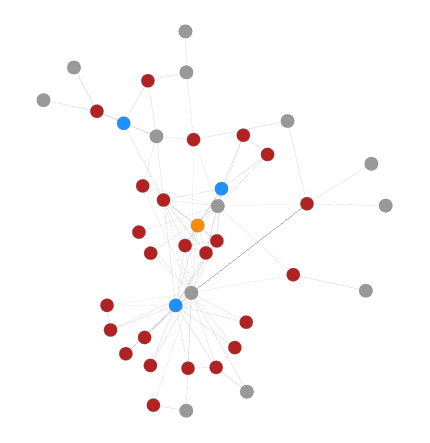}
                \caption{All nodes}
                \label{fig:npda}
        \end{subfigure}
       	\qquad 
		\begin{subfigure}[b]{0.2\textwidth}
                \centering
                \includegraphics[scale=1.0]{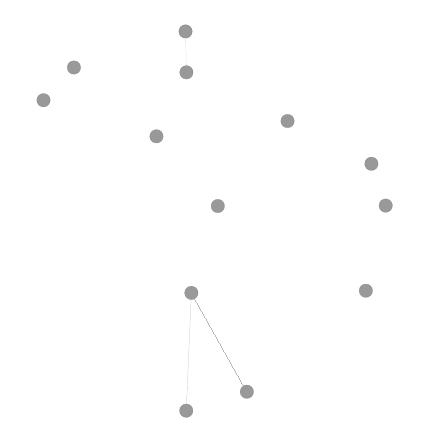}
                \caption{Twitter nodes}
                \label{fig:npdb}
        \end{subfigure}    
	\end{center}
	\caption{NPD community. Filtering results in central node loss and disconnection (grey=Twitter, blue=Facebook, orange=YouTube,
	red=other websites).}
	\label{fig:npdgraphs}
\end{figure}

We also present examples from the selected German language communities. The filtering of the non-electoral community in
Figure~\ref{fig:radgraphs} (58\% of nodes and 46\% of edges are retained) produces a similar effect to that of the EDL community, where
important central nodes such as the \emph{Besseres Hannover} website and those associated with other relevant information portals are
removed. Our last example in Figure~\ref{fig:npdgraphs} demonstrates the potential for severe loss of community structure when filtering is
applied (32\% of nodes and 3\% of edges are retained). In this example, although there is almost no communication (in terms of mentions and
retweets) between the Twitter nodes, they are considered members of the same community within the wider network.

\subsection{Community Characterization}

We now provide a characterization of the selected persistent dynamic communities shown in Figure~\ref{fig:timelines}, with a detailed
analysis of the member nodes; 
 in particular, those associated with external (non-Twitter) websites. 
Due to the sensitivity of the subject matter, and in the interests of
privacy, individual accounts and profiles from networks such as Twitter, Facebook and YouTube are not explicitly identified. Instead, we
restrict discussion to known extreme right groups and their affiliates. For each of the selected communities, we provide two alternative top
ten rankings of the website nodes; the first ranks the nodes in terms of their frequency of step membership, while the second ranks them in
terms of their total degree across all steps, normalized with respect to the total number of steps. In both cases, the membership ranking
distribution tends to be positively skewed, with a set of core community members assigned in the majority of steps, accompanied by a larger
set of peripheral members who were assigned intermittently.

\begin{figure}
	\begin{center}
        \begin{subfigure}[b]{0.4\textwidth}
       			\hskip -0.4em
                \centering
                \includegraphics[scale=0.49]{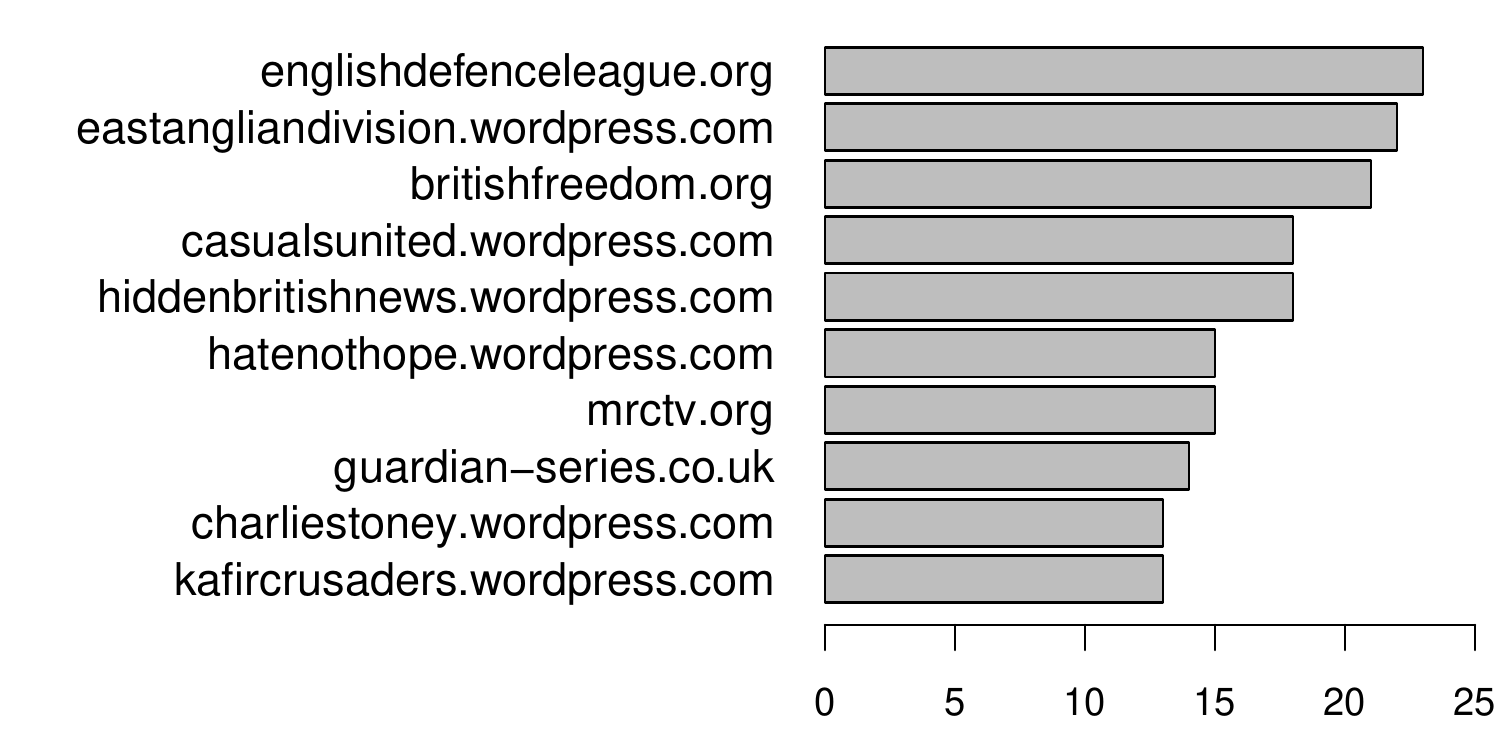}
                \caption{Membership frequency}
                \label{fig:edlfreq}
        \end{subfigure}
        \begin{subfigure}[b]{0.4\textwidth}
                \centering
                \includegraphics[scale=0.49]{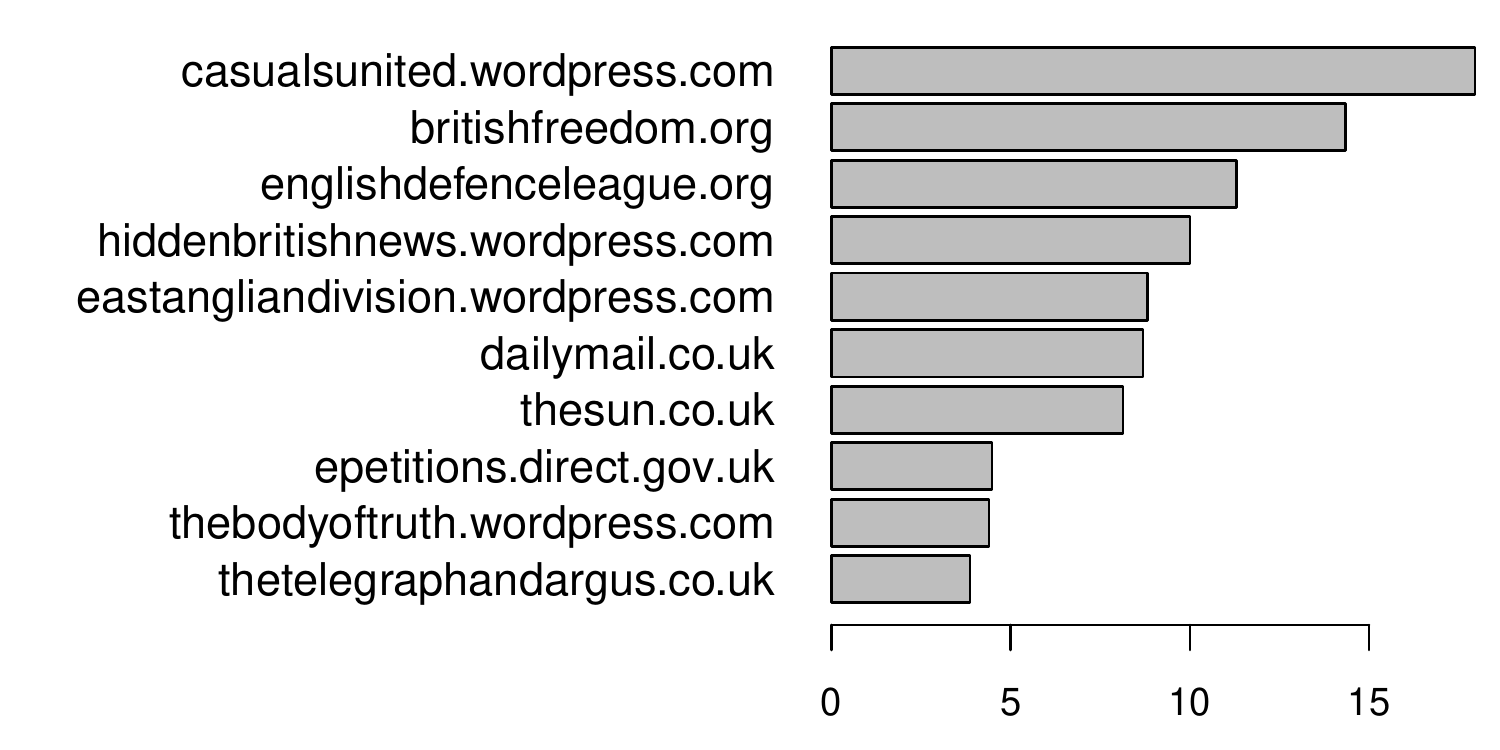}
                \caption{Normalized degree}
                \vskip -0.8em
                \label{fig:edldeg}
        \end{subfigure}
	\end{center}
	\caption{EDL community characterization, based on two alternative rankings of website nodes.}
	\label{fig:edlranking}
\end{figure}

Figure~\ref{fig:edlranking} presents the two rankings of website nodes for the EDL community. As might be expected, the frequency ranking
includes the official EDL and British Freedom websites. Other websites and blogs affiliated with the EDL are also present, including a
Casuals United blog. Of similar interest is a blog that appears to be a mocking reference to the established anti-extremist organization,
\emph{HOPE not hate}. The degree ranking produces broadly similar results, with notable exceptions being the appearance of mainstream media
websites. An analysis of the original URLs finds them to be associated with populist newspaper articles on topics such as the existence of
UK sex grooming gangs, Muslim integration, and immigration in general. Separate studies have found interest in these topics by other extreme
right groups \cite{ZwischenPropaganda2011}, and it would appear that mainstream media may unwittingly play a role in the occasional
promotion of material that coincides with extreme right ideology. Other articles that are interpreted as perceived media persecution of
groups such as the EDL are also promoted. Looking at the non-website nodes, official EDL Twitter and Facebook accounts can be observed,
along with those purporting to be affiliates. Similar YouTube channels are also present, including a British Freedom channel that was
suspended in late 2012 (this has since been replaced with a new channel according to their website).

\begin{figure}
	\begin{center}
		\begin{subfigure}[b]{0.4\textwidth}
                \centering
                \includegraphics[scale=0.49]{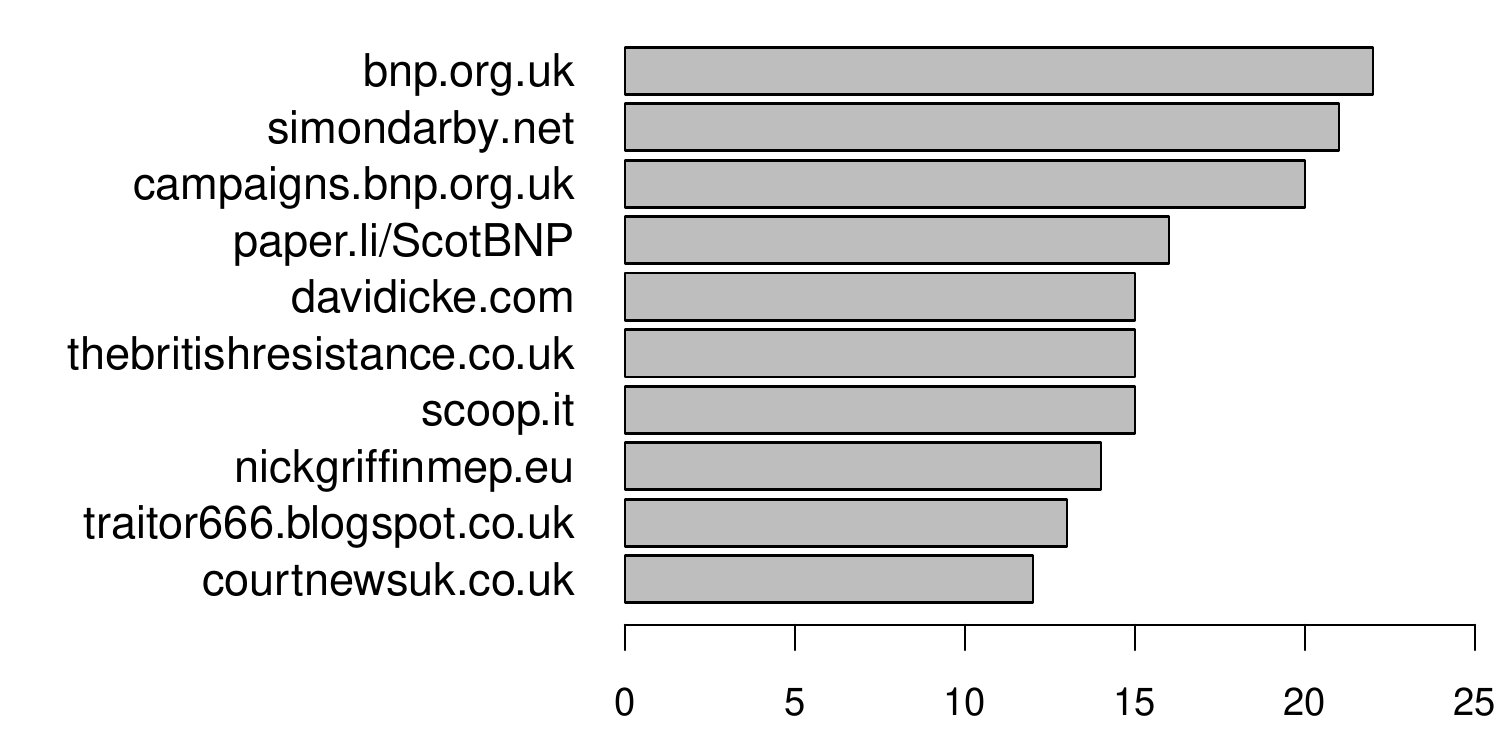}
                \caption{Membership frequency}
                \label{fig:bnpfreq}
        \end{subfigure}    
		\begin{subfigure}[b]{0.4\textwidth}
				\hskip -3.6em
                \centering
                \includegraphics[scale=0.49]{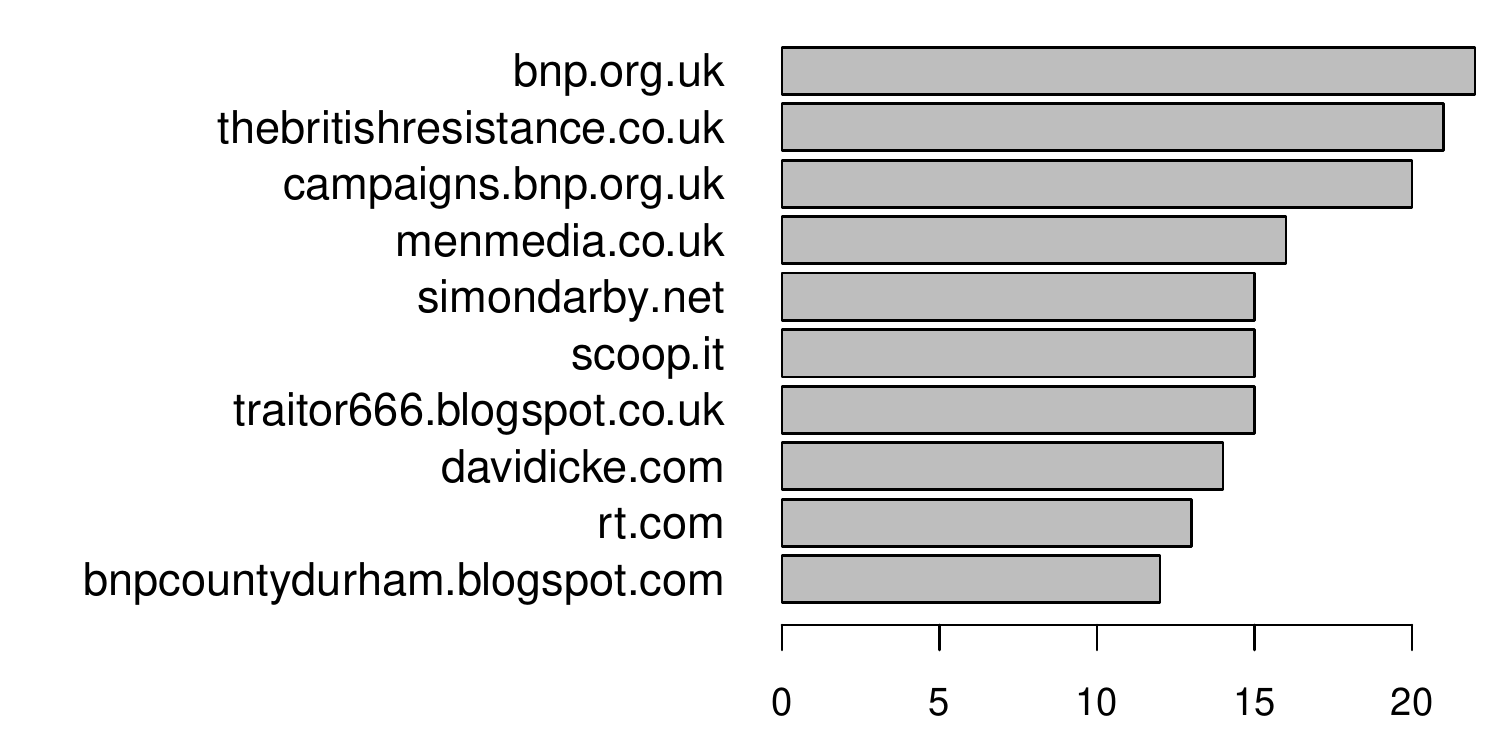}
                \caption{Normalized degree}
                \vskip -0.8em
                \label{fig:bnpdeg}
        \end{subfigure}    
	\end{center}
	\caption{BNP community characterization, based on two alternative rankings of website nodes.}
	\label{fig:bnpranking}
\end{figure}

We find similar membership composition in the BNP community, where the website rankings are presented in Figure~\ref{fig:bnpranking}. The
frequency ranking highlights websites associated with the BNP and that of its current leader, Nick Griffin. As mentioned in an earlier
section, this community also contains members associated with The British Resistance, a white nationalist group whose website banner
contains the 14-word slogan coined by the American white supremacist David Lane (\textit{``We must secure the existence of our people and a
future for White Children''}). As with the EDL community, mainstream media websites appear in the degree ranking, where the corresponding
article topics are similar to those promoted by the EDL. The presence of additional groups such as Combined ex-Forces (CxF) and the Infidels
can be found when the Twitter and Facebook nodes are inspected. The latter group originally splintered from the EDL, and although it has a
minor presence within the EDL community, it would seem that the BNP's efforts to court this group \cite{LowlesExtremisWhereNow2012} may
partly explain its more significant role in this community. The most prominent YouTube node is the official BNP channel, with other channels
of a nationalist nature appearing sporadically.

\begin{figure}[h!]
	\begin{center}
        \begin{subfigure}[b]{0.4\textwidth}
                \centering
                \hskip -0.3em
                \includegraphics[scale=0.49]{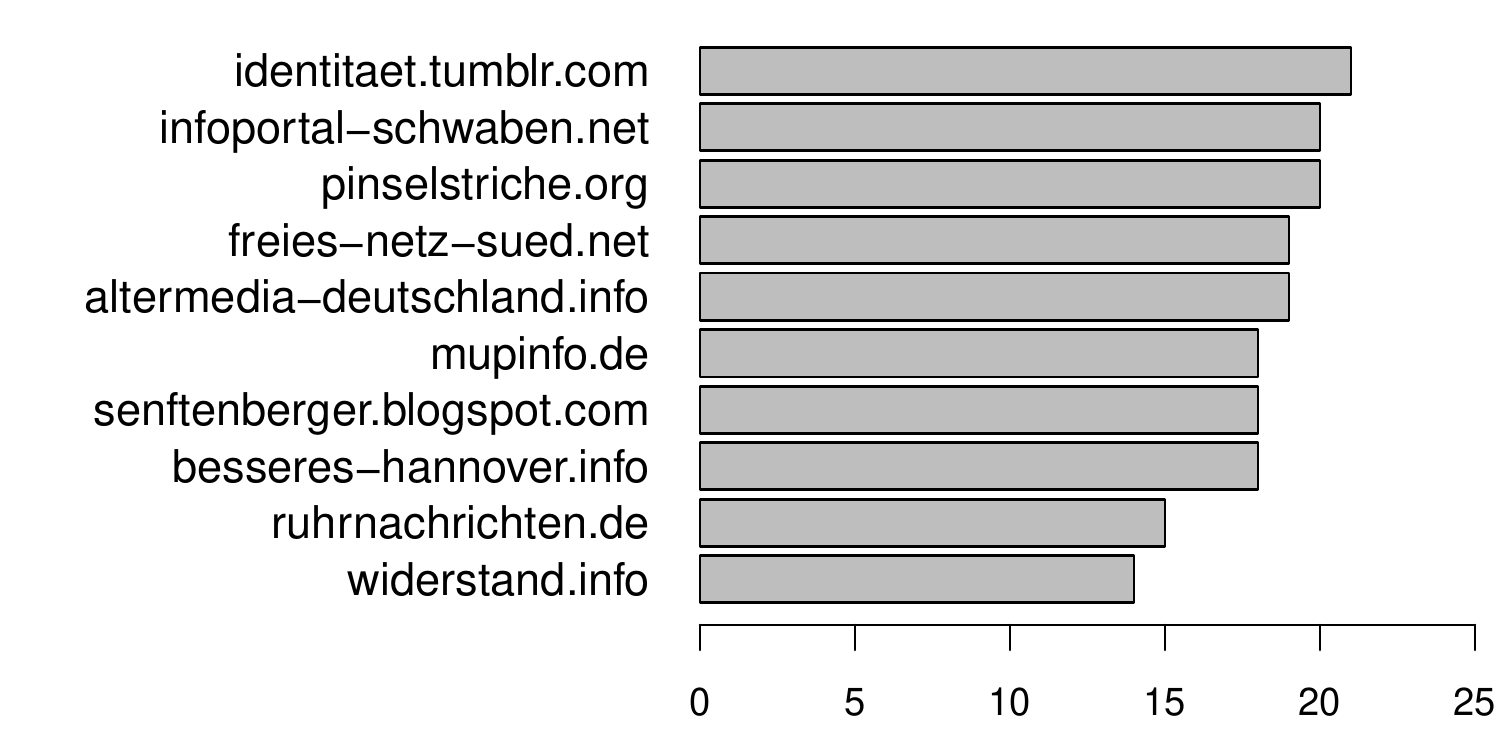}
                \caption{Membership frequency}
                \label{fig:radfreq}
        \end{subfigure}
        \begin{subfigure}[b]{0.4\textwidth}
                \centering
                \includegraphics[scale=0.49]{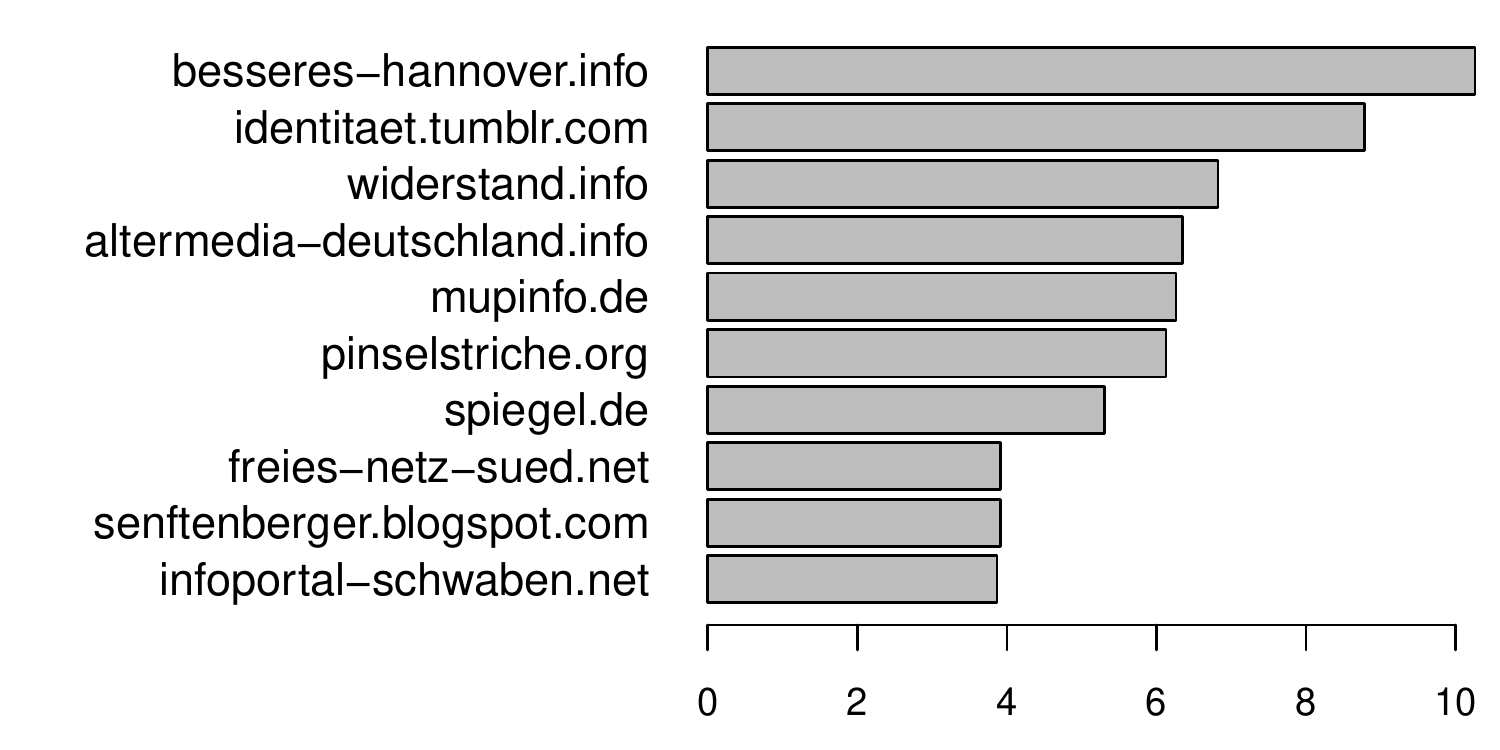}
                \caption{Normalized degree}
                \vskip -0.8em
                \label{fig:raddeg}
        \end{subfigure}
	\end{center}
	\caption{German non-electoral community characterization, based on two alternative rankings of website nodes.}
	\label{fig:radranking}
\end{figure}

\begin{figure}
	\begin{center}
		\begin{subfigure}[b]{0.4\textwidth}
                \hskip -0.4em
                \centering
                \includegraphics[scale=0.49]{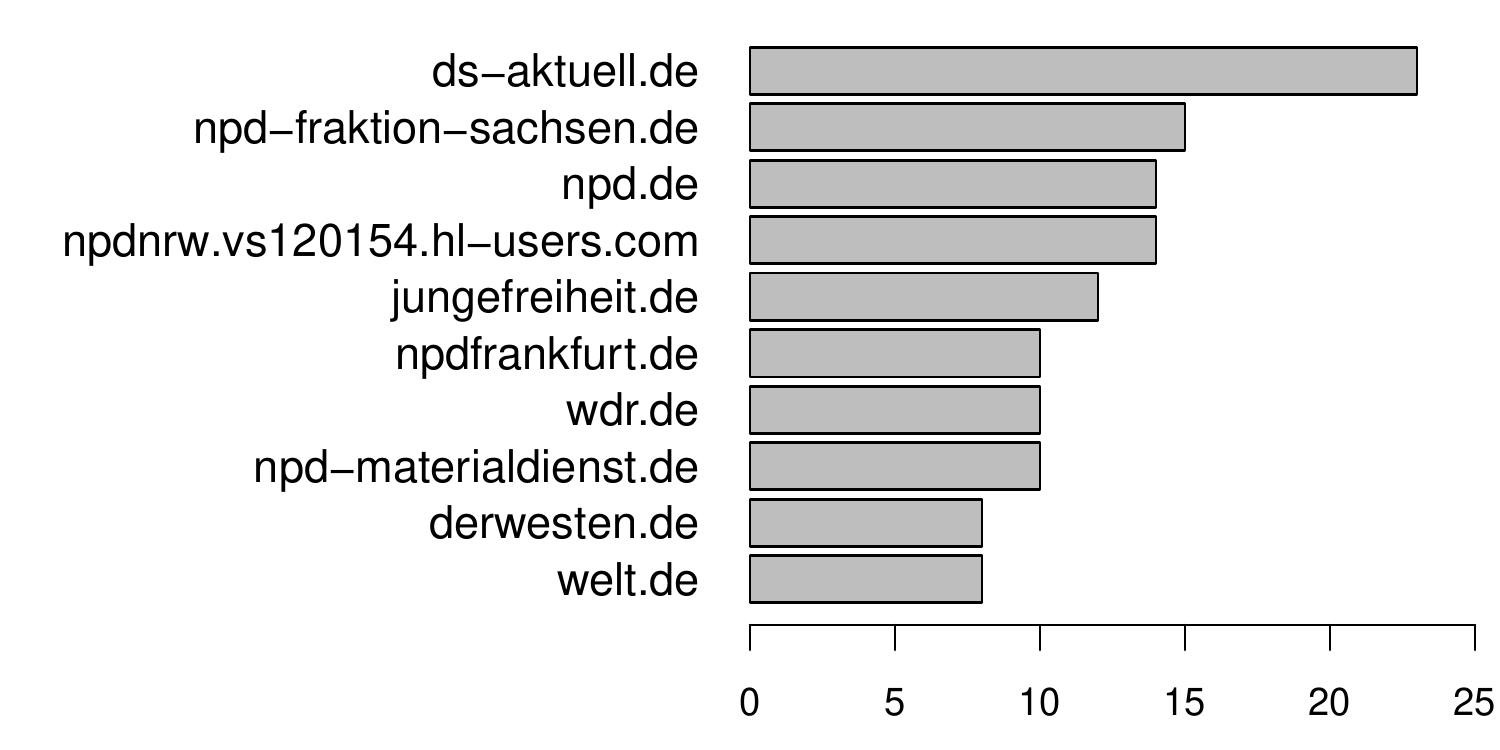}
                \caption{Membership frequency}
                \label{fig:npdfreq}
        \end{subfigure}    
		\begin{subfigure}[b]{0.4\textwidth}
                \centering
                \includegraphics[scale=0.49]{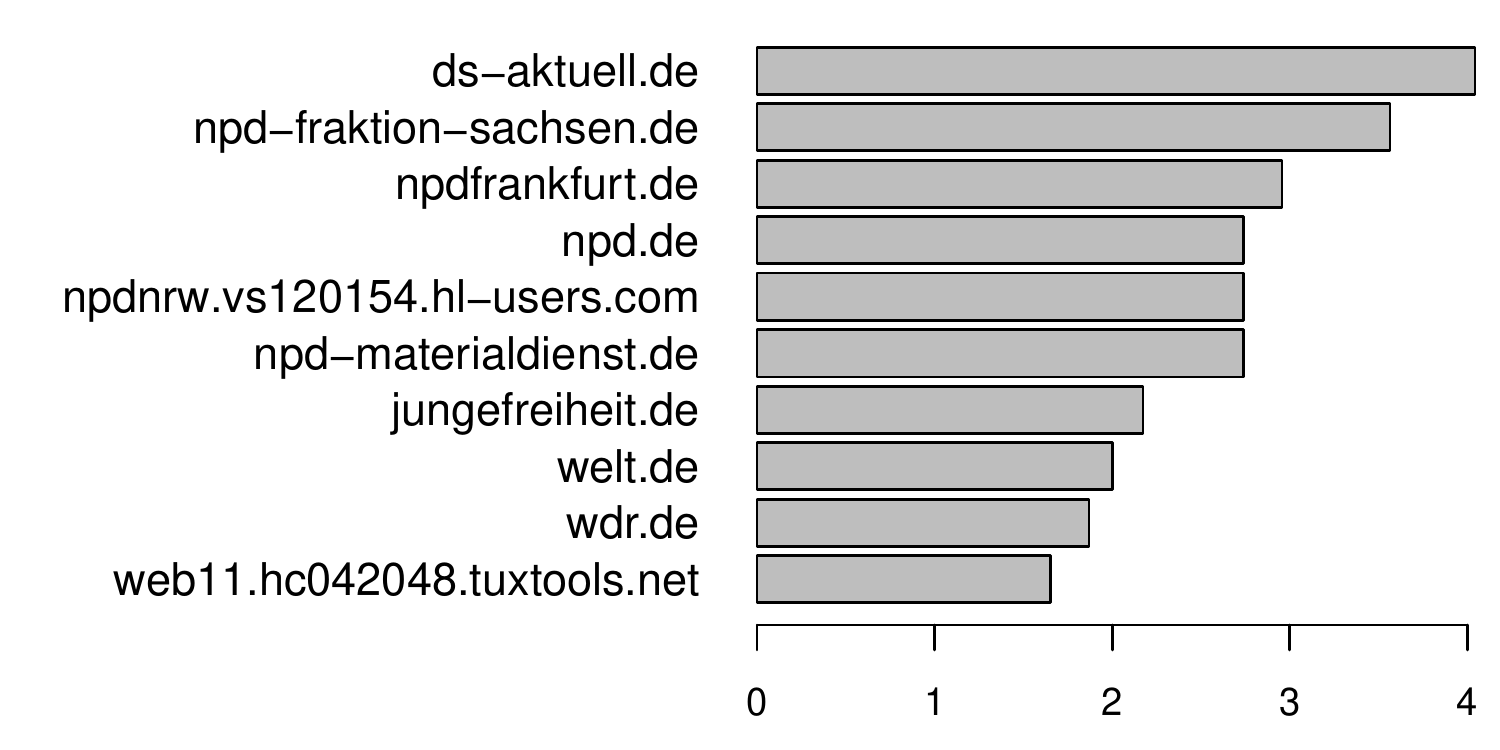}
                \caption{Normalized degree}
                \vskip -0.8em
                \label{fig:npddeg}
        \end{subfigure}    
	\end{center}
	\caption{NPD community characterization, based on two alternative rankings of website nodes.}
	\label{fig:npdranking}
\end{figure}

In the case of the German language website rankings, Figure~\ref{fig:radranking} shows blogs and websites from a variety of disparate
groups for the non-electoral community. The majority of these are associated with known non-electoral groups, although mainstream media
websites are also present. An analysis of the associated articles demonstrates once again the promotion of perceived persecution, for
example, the banning of the \emph{Spreelichter} and \emph{Besseres Hannover} groups features heavily here. We also see a connection to
electoral groups with the appearance of the NPD-affiliated MUPInfo website. Further connections to the NPD are visible as various NPD
Facebook page nodes are also intermittent members of this community. A notable temporary Facebook member node was a page (no longer
available) encouraging ``solidarity'' with Nadja Drygalla, a member of the 2012 German Olympic rowing team who left the tournament early due
to the neo-Nazi connections of her boyfriend, who was a former NPD election candidate \cite{ZeitDrygalla2012}. The website rankings for the
NPD community in Figure~\ref{fig:npdranking} are primarily composed of NPD-affiliated websites, with a minor number of mainstream websites
also appearing for the same reasons as before. Both communities feature a moderate YouTube presence, with nodes such as NPD channels, music
channels and individual channels featuring videos of demonstrations, including some of the \emph{Unsterblichen} (immortals) marches
that were orchestrated by \emph{Spreelichter} \cite{ZwischenPropaganda2011, ZeitSpreelichterBan2012}.

\subsection{Twitter Activity and External Events}

\begin{figure*}[!t]
	\begin{center}
        \begin{subfigure}[b]{1.0\textwidth}
                \centering
                \includegraphics[width=\linewidth]{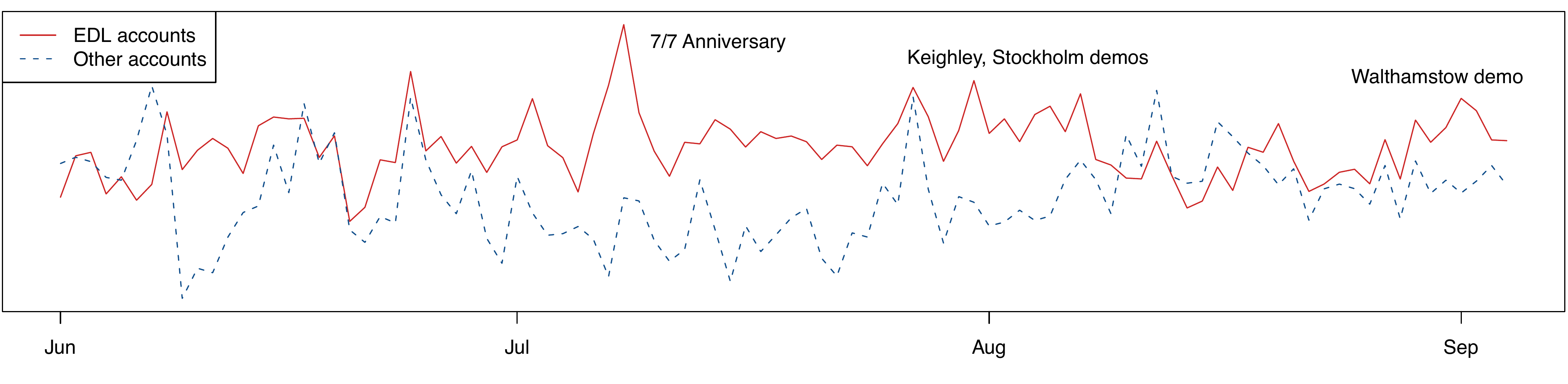}
                \vskip -0.4em
				\caption{EDL community tweet activity from June to September 2012.}
                \label{fig:edl_tweets}
        \end{subfigure}
		\vskip 0.4em
        \begin{subfigure}[b]{1.0\textwidth}
                \centering
                \includegraphics[width=\linewidth]{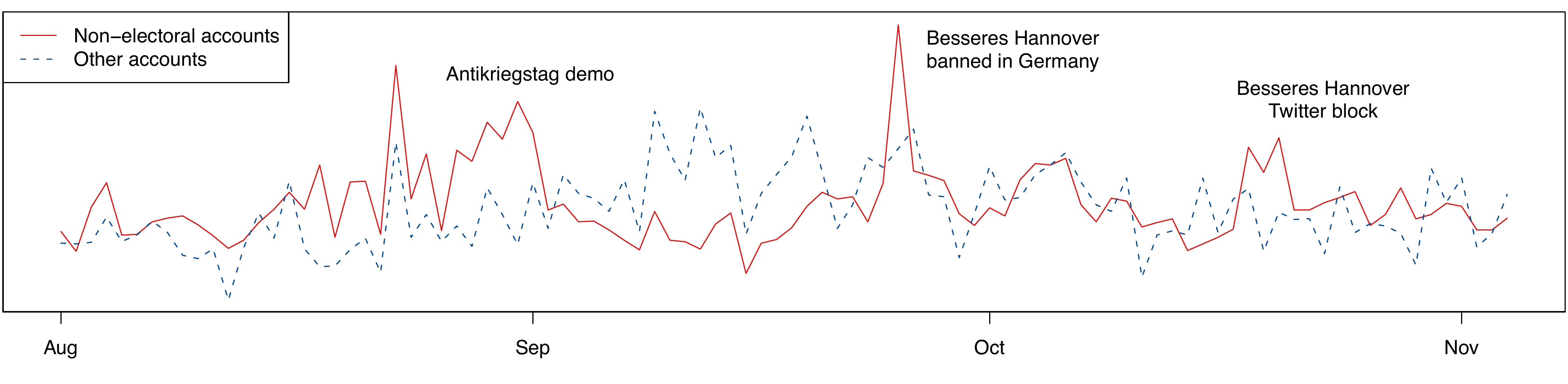}
                \vskip -0.4em
                \caption{German non-electoral community tweet activity from August to November 2012.}
                \label{fig:ne_tweets}
        \end{subfigure}
	\end{center}
	\vskip -0.6em
	\caption{Tweet activity plots of communities, produced via $z$-score normalization of tweet counts, for selected time periods.}
	\label{fig:tweets}
\end{figure*}

We also perform a brief inspection of community Twitter activity, providing examples for a single persistent dynamic community from both
data sets. Based on knowledge of external offline events associated with these communities, we generate three month plots of the daily total
tweet counts for (a) the accounts assigned in the corresponding two-week step community and (b) the remaining accounts in the data set,
where the raw counts are $z$-score normalized. In both cases, we find that peaks in tweeting activity by the community accounts may be
associated with external events. For example, the EDL activity plot in Figure~\ref{fig:edl_tweets} highlights increased activity around the
times of street demonstrations, with other events such as the anniversary of the July 7th, 2005 London bombings being of particular interest
to this community. Although the German non-electoral community plot in Figure~\ref{fig:ne_tweets} also contains raised activity for a
demonstration, other events such as the \emph{Besseres Hannover} ban (initially banned by the German authorities
\cite{ZeitBesseresHannoverBan2012}; their account was subsequently blocked by Twitter within Germany
\cite{GuardianBesseresHannoverTwitter2012}) have a similar impact. These peaks may introduce temporary community members, for example, a
file-sharing site containing an archive of the besseres-hannover.info website appears for a number of days at the time of the initial ban.

\subsection{Summary Discussion}
In general, the membership composition of the selected communities would appear to broadly correspond with contemporary knowledge of these
groups; for example, the distinction between the EDL and the BNP \cite{LowlesExtremisWhereNow2012}. To a certain extent, we also see
divisions according to the four-fold typology suggested by Goodwin~\etal \cite{GoodwinRadicalRight2012}, where these communities could be
characterized by three of the four types; organized political parties (BNP, NPD), grassroots social movements (EDL) and smaller groups
(German non-electoral). However, it could also be suggested that some overlap between these types can occur, for example, the presence of
the Infidels and British Resistance in the BNP community, or NPD-affiliated nodes in the German non-electoral community. In addition, the
finding of Bartlett~\etal \cite{BartlettDigitalPopulism2011} that online supporters of groups such as the EDL are more likely to demonstrate
than the national average may partly explain the increased levels of community Twitter activity at the time of external protests.
 
We also note the use of official Facebook profiles and YouTube channels by electoral parties and grassroots movements in addition to those
maintained by related individuals, while general blogging websites (often hosted in other countries) appear popular with most groups.
Separately, all groups appear to selectively reference mainstream media, directing traffic to material that coincides with associated
ideology. At this point, we also emphasize that the network representations used in this analysis do not provide coverage of all online
extreme right activity. As mentioned earlier, data collection from Twitter, YouTube and Facebook was restricted to publicly accessible
content. Given existing knowledge of the use of online platforms such as Facebook by the extreme right \cite{ZwischenPropaganda2011,
ISDPreventingCounteringExtremism2012}, a certain level of incompleteness is to be expected. In addition, the fact that we use Twitter 
to infer structure within the wider online network of the extreme right may also introduce incompleteness, where the network
representations are dependent on the initial identification of relevant profiles.

\section{Conclusions and Future Work}
\label{conclusions}

The online activity of extreme right groups has progressed from the use of dedicated websites to span multiple networks, including popular
social media platforms such as Twitter, Facebook and YouTube. As its role in the general dissemination of content has previously been
established, we have investigated the potential for Twitter to act as one possible gateway to communities located within this wider network.
By representing relations between the heterogeneous network entities with a single homogeneous network, we are able to identify extreme right
communities that would otherwise not be evident when considering a single online network alone. The use of heterogeneous data provides us
with a rich insight into the composition of the resulting communities. Our analysis has focused on the investigation of English and German
language communities using two separate data sets, where we have tracked community evolution in these inherently dynamic networks over an
extended period of time. Two persistent communities found in each of the data sets were selected for detailed analysis. We have discussed
the impact of heterogeneous nodes on community topology, and used these nodes to provide community characterizations, particularly in terms
of the extent to which these communities span multiple online platforms.

In our dynamic community analysis of both data sets, we have found that the individual step networks tended to exhibit a considerable level
of volatility. Although this may be due in part to data incompleteness, it is possible that such volatility may simply be a feature of the
online extreme right presence. It is likely that this corresponds to factors such as the continual emergence of new groups
\cite{MulhallExtremisBeginnersGuide2012}, while events such as the jailing of the EDL's Tommy Robinson
\cite{GuardianTommyRobinsonPassportJail2013} or a potential ban of the NPD \cite{ArzheimerExtremisNPDBan2013} are also likely to impact
volatility. We would like to address this issue in future work, which may involve an extension of the process currently used to track
dynamic communities. In addition, it would be worthwhile to investigate the potential for other social media platforms such as Facebook to
act as gateways to online extreme right activity, where a comparison could be made with the current results. This should also help to
address the issue of data incompleteness.

\section{Acknowledgments}

This research was supported by 2CENTRE, the EU funded Cybercrime Centres of Excellence Network and Science Foundation Ireland Grant 08/SRC/I1407 (Clique: Graph and Network Analysis Cluster).
%
%
%
%
%
\balance

\bibliographystyle{acm-sigchi}
\bibliography{extremists_websci2013}
\end{document}